\documentclass[prl,twocolumn,nofootinbib,preprintnumbers,superscriptaddress]{revtex4-1}

\usepackage{natbib}
\usepackage{amssymb,amsbsy,amsmath,amsfonts}
\usepackage{mathrsfs}
\usepackage{graphicx}% Include figure files
\usepackage{epsf,epsfig,float,latexsym,amsthm,fancyhdr,rotating}
\usepackage{graphics,psfrag,longtable}
\usepackage{slashed}
\usepackage{esint}
\usepackage{nicefrac}
\usepackage{braket}
\usepackage{mathtools}

\newcommand{\nn}{\nonumber}

\renewcommand{\eth}{\partial}
\newcommand{\be}{\begin{equation}}
\newcommand{\ee}{\end{equation}}

\usepackage{color}
\definecolor{orange}{rgb}{.9,.3,.0}

\usepackage[colorlinks,citecolor=blue,linktoc=all,linkcolor=cyan]{hyperref}

\begin{document}
%\preprint{MITP}
\title {%Toward resolving the debate on radiation forces and torques\\[1ex]
Superkicks and the photon angular and linear momentum density% controversy
}

%\author{If we like it, we will put our names here.}
\author{Andrei Afanasev}

\affiliation{Department of Physics,
The George Washington University, Washington, DC 20052, USA}

\author{Carl E. Carlson}

\affiliation{Physics Department, William \& Mary, Williamsburg, Virginia 23187, USA}

%\affiliation{And ... ?}

\author{Asmita Mukherjee}

\affiliation{Department of Physics, Indian Institute of Technology Bombay, Powai, Mumbai 400076, India}

% (Suggestion for abstract : Asmita) We address a problem of proper definition of momentum density for a spatially structured electromagnetic field. We show that the expressions for the momentum and angular momentum  obtained locally are not the same when one uses the canonical energy-momentum tensor instead of the symmetric Belinfante  energy-momentum tensor in QED. This has important consequences for interaction of matter with structured light, for example, twisted photons; and would give drastically different results for forces and angular momenta induced on small test objects.  We show, with numerical estimates of the size of the effects, situations where the canonical and symmetrized forms induce very different torques and recoil momenta on small objects or atomic rotors, over a broad range of circumstances.

\begin{abstract}

We address the problem of determining the physically correct
definition for the momentum and angular momentum densities in a
spatially structured electromagnetic field, given that the expressions
are not the same when one uses
%We address a problem of proper definition of momentum density for spatially structured %electromagnetic fields.
%We show that the expressions for the momentum and angular momentum  obtained locally are not the same when one uses 
the canonical energy-momentum tensor instead of the symmetric Belinfante  energy-momentum tensor in  electrodynamics. This has important consequences for interaction of matter with structured light, for example, twisted photons; and would give drastically different results for forces and angular momenta induced on small test objects.  We show, with numerical estimates of the size of the effects, situations where the canonical and symmetrized forms induce very different torques or (superkick) recoil momenta on small objects or atomic rotors, over a broad range of circumstances.
%Canonical expressions for the energy-momentum tensor are obtained from symmetry transformations on the Lagrangian using the Noether construction.  The result is not generally symmetric in its two indices, but can be symmetrized by adding  total derivative.  The subsequent expressions for the momentum and angular momentum are the same when integrated over all space, if the boundary conditions are suitable, but are not in general locally the same.  In particular, they are not the same for structured light, for example, for twisted photons.  Hence they can predict different results for forces and angular momenta induced on small test objects.  We will show, with numerical estimates of the size of the effects, situations where the canonical and symmetrized forms induce very different torques on small objects, over a broad range of circumstances.  We will also comment on some remarkable, albeit sensitive to details, cases where the predicted radiation forces on small objects are very different based on momentum densities obtained from the canonical and symmetrized cases.  
\end{abstract}
\date{\today
}
\maketitle

%%%%%%%%%%%%%%%%%%%%%%%%%%%%%%%%%%

\paragraph{Introduction.}

%%%%%%%%%%%%%%%%%%%%%%%%%%%%%%%%%%

There remains disagreement over the correct expression for the linear or angular momentum density of electromagnetic beams.  Light beams with non-trivial wavefronts---structured light---and in particular twisted photon beams, give opportunity both for adjudicating the controversy and for dramatic results whatever the outcome. 

Twisted photons are vortex-like solutions to the Helmholtz equation in cylindrical coordinates which have fields swirling around a vortex line, and whose angular momentum in the direction of propagation, $m_\gamma$, is any integer times $\hbar$.  They stand in stark contrast to plane wave photons where $m_\gamma = \pm \hbar$ only.  For reviews, see~\cite{2011AdOP....3..161Y,Bliokh:2015doa}, and for further discussions of momentum definitions, see~\cite{Bliokh:2015doa,2013EJPh...34.1337B,2009PhRvL.102k3602A,1978OptCo..24..185H,2019PhRvA..99f3832W,2014OExpr..22.6586O,2002PhRvL..88e3601O,2003PhRvL..91i3602G}.

Experimentally, the large total angular momentum is verified. For example in~\cite{1995PhRvL..75..826H,1996PhRvA..54.1593F},  where twisted photons were absorbed by small objects suspended in a viscous fluid, the objects were observed to acquire spins in agreement with the heightened angular momentum of the twisted photon.  However, in these experiments the objects, although small, are  large enough to absorb the entire twisted beam and so measured total angular momentum and not the local angular momentum density of the beam.

Quantum-state control of trapped ions using laser beams is presently one of the most promising techniques for quantum computing \cite{PhysRevLett.74.4091}. The spatial extent of the ion's wave function in a harmonic oscillator trap may be, for example, about 5 nm \cite{2016NatCo...712998S} if the ion is cooled close to oscillator's  ground state. With such localization of the objects compared to $\mu$m wavelength of light used to manipulate them, the question of how one should calculate the linear and angular momentum density of electromagnetic field  becomes important.  There is a canonical procedure that leads to a certain expression, reviewed below, that can in turn be obtained from a canonical expression for the energy-momentum tensor that is not symmetric in its two indices.  Citing both aesthetics and needs of General Relativity, one can add a total derivative to make the energy-momentum tensor symmetric, a procedure pioneered by Belinfante~\cite{1940Phy.....7..449B} and by Rosenfeld~\cite{rosenfeld1940energy}, and then obtain a different expression for the angular momentum density, also reviewed below.  Because of the total derivative, the integrals that give the total angular momentum are identical, if the surface terms cause no problem.

However, as emphasized particularly in~\cite{Leader:2015vwa,Leader:2017htb}, when structured light shines on rings, or generally on small objects with open centers, the angular momentum absorbed or the torque induced depends on the angular momentum density expressions at radii where the matter exists, and the expressions are rather different for the canonical and symmetrized or Belinfante cases.   Further, Ref.~\cite{2019PhRvA..99f3832W} suggests an alternative geometric spin Hall effect in light, where the canonical and Belinfante predictions are quite distinct, so that measurements could show that either or both of them must be wrong.
Ref.~\cite{2014OExpr..22.6586O} showed that one could obtain the same momentum in a confined volume using the canonical density as would be gotten using the Belinfante density if one accounted for the surface terms, but did not, to our minds, argue decisively which density was physically correct.
Refs.~\cite{2002PhRvL..88e3601O,2003PhRvL..91i3602G} report measurements of transverse optical forces on small particles in Bessel beams, however at selected locations where the canonical and Belinfante force predictions happened to be the same.   We will comment on further opportunities in this type of measurement in the text below.

Alternatively, one may discuss the effect of twisted light on small objects in terms of superkicks, to use a term coined in~\cite{2013JOpt...15l5701B}, to describe the effects of the sometimes quite large azimuthal components of the linear momentum density. Specific examples of superkicks and possibilities to observe this quantum effect were recently considered in Refs.\cite{Afanasev_21,Ivanov_22}.  %  The swirling fields of the twisted photon give a linear momentum density that has transverse, specifically azimuthal, components as well as longitudinal (along the propagation direction) components.  The transverse components can be, relative to the local photon energy density, very large as one approaches the vortex axis of the twisted photon.  A small enough test object located on one side of but near to the vertex axis can receive a significant transverse kick, or superkick.  
The calculated size of the superkick depends critically upon whether one uses the canonical or Belinfante expression.  Close to the vortex line, the canonically calculated superkick is much larger, and farther out %where the canonical and Belinfante results are similar in magnitude, 
there are broad regions where they differ in sign.

There is a separable discussion, which we will not enter, in both the canonical and symmetrized contexts, of how %, and even if it is possible, in the fully interacting case 
to write the angular momenta for the fermions and vector bosons in QED or QCD, with distinct spin and orbital angular momentum for each field, while maintaining gauge invariance.  For a review of this discussion, see~\cite{Leader:2013jra}. %This separable discussion we will not enter here.

The goal of this paper %letter 
is to show examples, with numerical estimates, for the forces, torques, and accelerations, of situations where the canonical and Belinfante form of the linear or angular momentum density lead to very different results.  We will begin with a short review, followed by studies of twisted photons axially striking hollow cylinders, of twisted photons impinging on and accelerating a two-ion rotor, of small particles struck while off-axis in a twisted beam, and of radiation pressure on small objects appearing as tractor beams in limited regions.

%%%%%%%%%%%%%%%%%%%%%%%%%%%%%%%%%%

%\section{Selected formulas and results}

%%%%%%%%%%%%%%%%%%%%%%%%%%%%%%%%%%

%%%%%%%%%%%%%%%%%%%%%%%%%%%%%%%%%%

\paragraph{A brief review of the formalism :}

%%%%%%%%%%%%%%%%%%%%%%%%%%%%%%%%%%

%It is suitable to begin with a short summary of the different expressions for the local momentum and angular momentum of photons.  
The electromagnetic Lagrangian is
$
L = - %\frac{1}{4 \mu_0 } 
F_{\alpha\beta} F^{\alpha\beta}/(4\mu_0)	.
$
Studying the response of the Lagrangian to coordinate translations leads to a canonical and conserved (in the first index) energy-momentum tensor (see, e.g.,~\cite{Jauch:1976ava,Bjorken:1965zz}) 
\begin{align}
T^{\mu\nu} =	- \frac{1}{\mu_0} F^{\mu\alpha} \eth^\nu \! A_\alpha - g^{\mu\nu} L	.
\end{align}
The tensor is not symmetric.  It can be made symmetric by adding a total derivative 
$ - \eth_\alpha \left( F^{\mu\alpha} A^\nu \right)/ \mu_0$, which by virtue of the equations of motions leads to a symmetric or Belinfante energy-momentum tensor
\begin{align}
\theta^{\mu\nu} = - \frac{1}{\mu_0} F^{\mu\alpha} F^\nu_{\ \alpha} -	g^{\mu\nu} L	.
\end{align}
The tensor remains conserved.

The linear momentum densities $\mathcal P^\nu$ are the $T^{0\nu}\!/c$ or $\theta^{0\nu}\!/c$ components of these tensors, so that
\begin{align}
\vec {\mathcal P} = \left\{
			\begin{array}{ll}
			 \epsilon_0 \vec E \cdot (\vec\nabla) \vec A	\,,	&	\quad \text{canonical},	\\
			 \epsilon_0 \vec E \times \vec B		\,,	&	\quad	\text{symmetric or Belinfante}.
			\end{array}
\right.
\end{align}
The last is also the Poynting vector times $1/c^2$, and the notation in the first line means 
$\mathcal P^i_\text{can} = \epsilon_0 \sum_{j=1}^3 \ E^j  \nabla_i A^j$.  

Further, Lorentz and rotation transformations lead to a canonical angular momentum tensor
\begin{align}
\mathcal M^{\alpha\mu\nu} = x^\mu T^{\alpha\nu} - x^\nu T^{\alpha\mu} 
	+ \frac{ \eth L}{\eth(\eth_\alpha A_\beta) } \Sigma^{\mu\nu}_{\beta\gamma} A^\gamma	,
\end{align}
with $\Sigma^{\mu\nu}_{\beta\gamma} = g^\mu_\beta g^\nu_\gamma - g^\mu_\gamma g^\nu_\beta	\,.$   The canonical angular momentum densities $\vec {\mathcal J}$ come from $\mathcal M^{0ij}/c$.   For the symmetrical or Belinfante case, one just takes $\vec r$ times the corresponding momentum density, with no explicit spin term.  
\begin{align}
\vec {\mathcal J} = \left\{
			\begin{array}{l}
	\epsilon_0 \vec E \cdot (\vec r \times \vec\nabla) \vec A	+ \epsilon_0 \vec E \times \vec A
		\,,		\quad \text{canonical},	\\
			 \epsilon_0 \, \vec r \times ( \vec E \times \vec B )		\,,		
			 	\qquad	\text{symm. or Belinfante}.		
			\end{array}
\right.
\end{align}
The two expressions differ by a total derivative.  But they differ locally, so do not lead to the same torque upon small test objects.

For the $\mathcal J_z$ components, the differences when considering structured light are large and robust.  As the discussion proceeds, we will begin with these components. %, and later discuss radiation pressure or forces induced along the propagation direction. %, where the differences between the canonical and Belinfante predictions can also be dramatic but sensitive to detail.   

In the paraxial approximation, the transverse part of the vector potential % for either Laguerre-Gauss or Bessel-Gauss twisted photons 
is
\begin{align}	\label{eq:paraxial}
\vec A(\vec r, t) =  \hat \epsilon \, u(\rho, \phi,z) e^{i(k z-\omega t)}		.
\end{align}
The $\rho, \phi,z$ are cylindrical coordinates; $z$ is the overall propagation direction of the beam; and 
$\hat\epsilon$ is a polarization vector
\begin{align}
\hat \epsilon = a \hat\eta_+ + b \hat\eta_- \,,
\end{align}
with $\eta_\Lambda = ( -\Lambda \hat x - i \hat y ) / \sqrt{2}$ and $|a|^2+|b|^2 =1$, with $\Lambda = \pm 1$.  Also we will let 
$\sigma_z \equiv |a|^2 - |b|^2$.  
Using pointed brackets to denote the time average, the $z$-components of the angular momentum densities are
\begin{align}
\braket{\mathcal J_z}_\text{can} &= \left( \frac{1}{2} \epsilon_0 \omega \right)  
	\left( \ell + \sigma_z \right) |u|^2,
		\nn\\
\braket{\mathcal J_z}_\text{Bel} &= \left( \frac{1}{2} \epsilon_0 \omega \right)  
	\left[		\left( \ell + \sigma_z \right) |u|^2 
		- \frac{\sigma_z}{ 2 \rho } \frac{\eth (\rho^2 |u|^2) }{ \eth \rho }  	\right]	.
\end{align}
The total angular momentum along the beam direction is 
$m_\gamma \hbar = ( \ell + \sigma_z ) \hbar$, on a per photon basis.

 We will work with Bessel-Gauss solutions for the function $u$;  the results are similar to Laguerre-Gauss beams for suitable choices of parameters.  For the Bessel-Gauss beams%, $u$ has no $z$ dependence and
\begin{align}
u(\rho,\phi,z) = u(\rho,\phi) =  A_0  J_\ell(\kappa \rho) e^{i \ell\phi}   e^{-\rho^2/w_0^2}		.
\end{align}
The monochromatic angular frequency is $\omega$, $k=\omega/c$, $\kappa = k \sin\theta_k$, with $\theta_k$ the pitch angle whose smallness defines the paraxial approximation,  and $J_\ell$ is a Bessel function.  The Gaussian width of  the envelope is $w_0$. %, which can generally be taken much larger than the photon's wavelength.  

%%%%%%%%
\begin{figure}[t]
\begin{center}

\centerline{ (a) } \smallskip
\includegraphics[width = 0.98 \columnwidth]{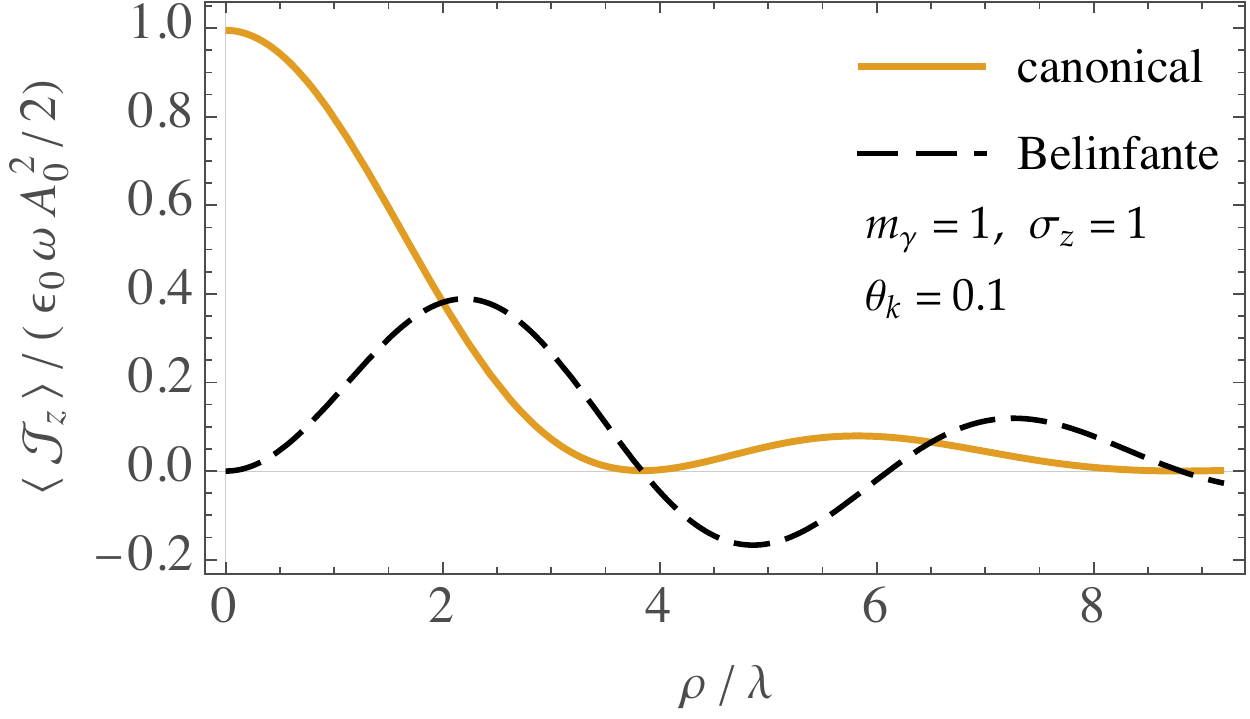}

\vskip 2 mm

\centerline{ (b) } \smallskip
\includegraphics[width = 0.98 \columnwidth]{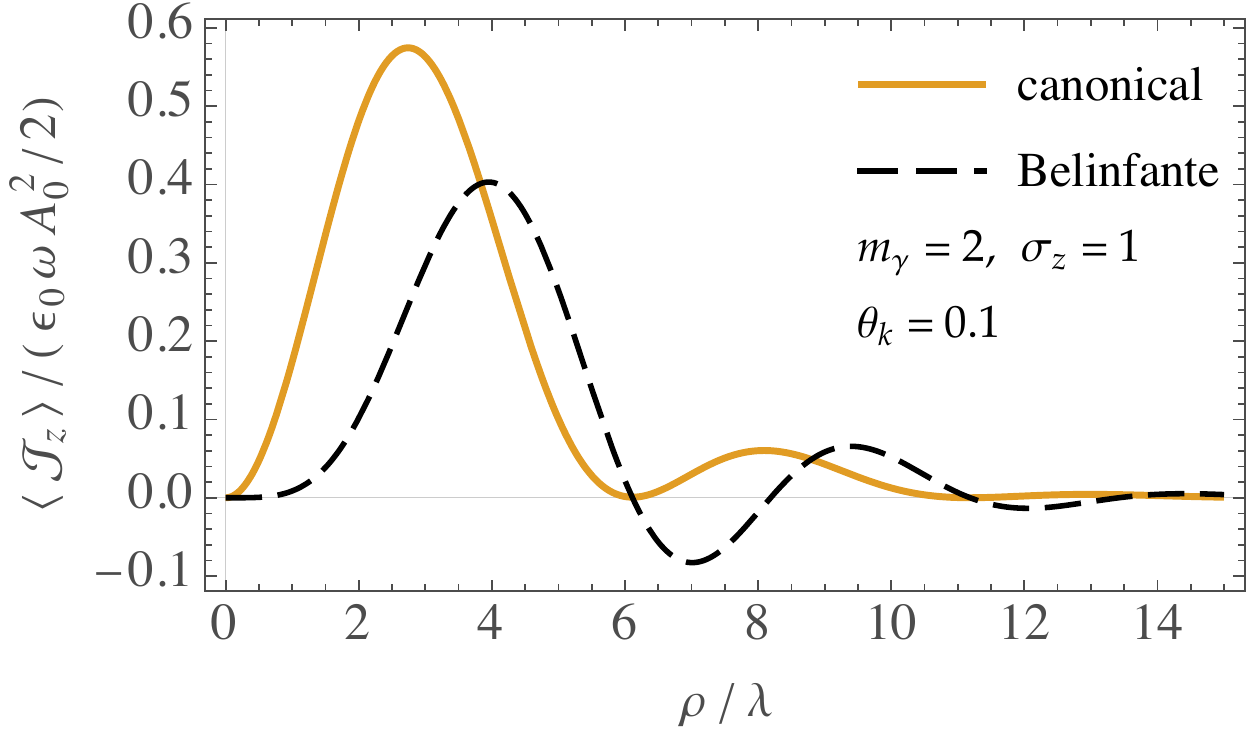}

\caption{Angular momentum density on a ring of radius $\rho$ for a twisted light beam of total angular momentum $m_\gamma =1$ (upper panel) or  $m_\gamma =2$ (lower panel)  and circular polarization $\sigma_z =1$, with angular frequency $\omega$ and $A_0$ normalizing the strength of the beam's electric field.   %The integrated angular momentum is the same for the canonical and Belinfante cases. %, and even in detail, integrating $\braket{\mathcal J_z}$ over the area between two near-zeros of the canonical case---or between the origin and the first near-zero---gives closely the same result for the two cases.  
The Belinfante case has regions where the angular momentum density swirls in a direction opposite to the overall angular momentum. 
}
\label{fig:angular2}
\end{center}
\end{figure}
%%%%%%%%

Plots of these densities as a function of distance from the vortex line, given in terms of the wavelength $\lambda$, are shown in Fig.~\ref{fig:angular2} for total angular momenta $m_\gamma = 1$ and $2$, $\theta_k=0.1$; polarization $\sigma_z=1$, selected pitch angle, and Gaussian envelope width $w_0 = 10 \lambda$.  The canonical angular momentum density is never negative (and with no paraxial approximation is never zero except at $\rho=0$).  However, the symmetric or Belinfante case has regions where the angular momentum density swirls in a direction opposite to the overall angular momentum.  These remarkable opposite swirling regions are broad and the predictions of their location and strength are not sensitive to making or not making the paraxial or other approximations.  

%%%%%%%%%%%%%%%%%%%%%%%%%%%%%%%%%%

\paragraph{Twisted photons incident on a hollow cylinder.}

%%%%%%%%%%%%%%%%%%%%%%%%%%%%%%%%%%

\begin{figure}[b]
    \centering
    \includegraphics[width = 0.63 \columnwidth]
    {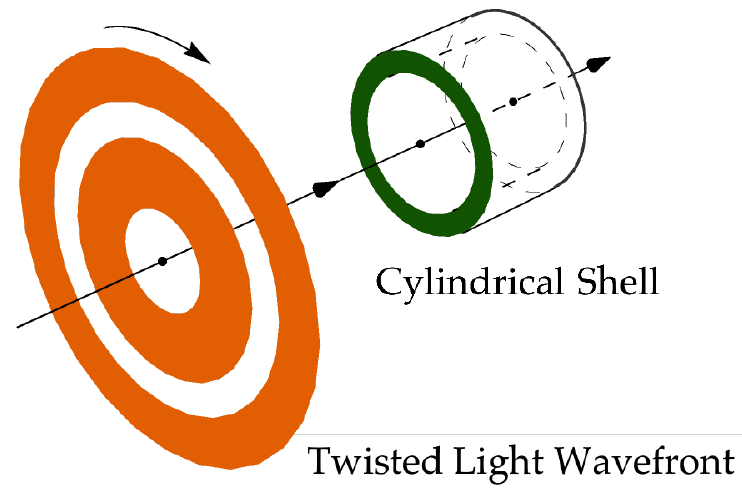}   \hfil 
    \raisebox{6 mm}{\includegraphics[width = 0.32
    \columnwidth]
    {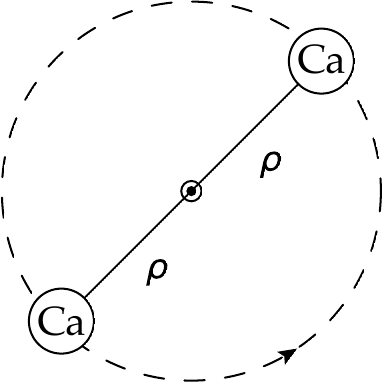}  }
    \vskip 2 mm
    
    \hskip 5 mm   (a)  \hskip 50 mm (b)
        
    \caption{(a) Twisted light hitting a hollow cylinder, with axes coincident, (b) A two-ion Calcium rotor.}
    \label{fig:twcylinder}
\end{figure}

To see what torques and angular velocities might be imparted to real test objects, we consider a specific measurement situation.  We think of the twisted photon shining on a ring, or in three-dimensions a test object which is a hollow cylinder with its axis identical to the vortex axis of the twisted photon beam, as depicted in Fig.~\ref{fig:twcylinder}(a). We will suppose that all the light, and the angular momentum it contains, hitting the front edge of the cylinder is absorbed by the cylinder.   We calculate first the angular acceleration the cylinder would have if it were in free space, and then calculate the terminal angular velocity it would obtain if it be suspended in a viscous fluid.  

%%%%%%%%%%%%%%%%%%
%%%%%%%%%%%%%%%%%%

If the hollow cylinder has a average radius $\rho$ with inner and outer radii $\rho \pm (1/2) \Delta \rho$, then the torque from absorbing the light on the front face of area $\sigma$ is
\begin{align}
\braket{ \tau_z } = \sigma c   \braket{ \mathcal J_z }	= 	( 2 \pi \rho \Delta\rho)  c \braket{ \mathcal J_z }.
\label{tau}
\end{align}
%where $\sigma$ is the cross section.

We obtain the normalization $A_0$ from the total power in the Bessel-Gauss beam, which we get by integrating the $z$-component of the Poynting vector.  (The latter gives the energy flux whether we use the canonical or Belinfante energy-momentum tensor.)   The $z$-component of the Poynting vector becomes, paraxially,
\begin{align}
\braket{S_z} &= \frac{ \omega k A_0^{\,2} }{2 \mu_0}  J^2_{m_\gamma-\Lambda}(\kappa \rho)	\,
		\exp (- 2 \rho^2/w_0^2)	,
\end{align}
for a polarized situation where 
$\sigma_z = \pm 1 = \Lambda$.
The beam's time average power is $\braket{ P}$, and we obtain the normalization $A_0$ from
\begin{align}
\braket{ P } &= \int_0^\infty   \braket{S_z}   2\pi \rho d\rho	%	\nn\\	
%&= \pi \omega^2 A_0^{\,2}  \int_0^\infty  J^2_{m_\gamma-\Lambda}(\kappa \rho) 
%	\exp (- 2 \rho^2/w_0^2)	 \rho \, d\rho		
.
\end{align}
We use $\braket{ P } = 4 \,$mW, wavelength $\lambda = 729$ nm, and $w_0 = 10 \lambda$. 
The latter two numbers match conditions in~\cite{2016NatCo...712998S,2018NJPh...20b3032A}, and the first matches the quoted power of a twisted beam delivered on a target in~\cite{1995PhRvL..75..826H}.  For definiteness, we consider the $m_\gamma = 2$ case,  wall thickness $\Delta\rho = 0.5 \, \mu$m, length $L= 2 \,\mu$m. We will give explicit numbers for $\rho =2 \,\mu$m, a value of $\rho$ near the peak of the angular momentum density for the $m_\gamma=2$ canonical case; see Fig~\ref{fig:angular2}(b).   Results for other values of $\rho$ can be scaled from the results in Fig~\ref{fig:angular2}.

The moment of inertia is
$
I = M \rho^2 = 2\pi  \rho_m \,  \rho^3 \Delta\rho L,
$
where $M$ is the mass of the cylindrical shell and $\rho_m$ is its mass density, which for the sake of illustration  we take as twice the density of water. The  angular acceleration $\braket{\alpha } = \braket{ \tau_z } / I$ for the cylinder in free space is
\be
\braket{\alpha }  \approx  \left\{
\begin{array}{ll}
5.5 \times 10^6 \text{\,rad/s}^2	&
\quad \text{canonical} , \\
2.3 \times 10^6 \text{\,rad/s}^2	&
\quad \text{Belinfante},
\end{array}
\right.
\ee
where $\tau_z$ is given in terms of $J_z$ in Eq.(\ref{tau}).  

If the cylinder is in a viscous medium, there is a drag torque on it,
$
\tau_\text{drag} = - 4\pi \eta \rho^2 \ell \, \Omega		,
$
where $\eta$ is the viscosity and $\Omega$ is the cylinder's angular rotation frequency. If the medium is kerosine ($\eta = 1.64 \times 10^{-3} \text{\,N} \cdot  \text{s/m}^2$), the terminal rotation frequency is       
\be
f  \approx  \left\{
\begin{array}{ll}
0.55   \text{\,Hz}	&
\quad \text{canonical} , \\
0.23   \text{\,Hz}	&
\quad \text{Belinfante}.
\end{array}
\right.
\ee
Again, this is the prediction for one radius. Using other radii will give different results following Fig.~\ref{fig:angular2}. Note that the $0.5 \,\mu$m shell thickness is narrow enough to fit within the negative region of the Belinfante curves in that Figure. 

%%%%%%%%%%%%%%%%%%%%%%%%%%%%%%%%%%

\paragraph{A two-ion rotor.}

%%%%%%%%%%%%%%%%%%%%%%%%%%%%%%%%%%

Another situation distinguishing the canonical and Belinfante calculations is twisted photons striking a two-ion rotor.  Our discussion is inspired by the working rotor described in~\cite{2019PhRvL.123m3202U}.

We shall describe the mechanism in term of superkicks.  These come from the azimuthal component of the momentum density, which paraxially is
\begin{align}
\braket{\mathcal P_{\phi}}_\text{can}
&= \frac{\epsilon_0 \omega  A_0^{\,2} }{2 \rho} 
(m_\gamma - \Lambda)  J_{m_\gamma-\Lambda}^2(\kappa \rho) 	,	\nn\\
\braket{\mathcal P_{\phi}}_\text{Bel} &= \frac{ \epsilon_0\omega A_0^{\,2} }{2 }  \,  
	\kappa \, J_{m_\gamma}(\kappa \rho)  J_{m_\gamma-\Lambda}(\kappa \rho)	.
\end{align}
The photon number density in either case is
\begin{align}
\braket{n_\gamma} &= \frac{\epsilon_0 \omega A_0^{\,2} }{2 \hbar}
	 J_{m_\gamma-\Lambda}^2(\kappa \rho)	.
\end{align}
The transverse momentum kick or superkick  at distance $\rho$ from the vortex line is then
\begin{align}
\braket{ p_{\phi} } = \frac{  \braket{ \mathcal P_{\phi} }  } {  \braket{n_\gamma}  }
	= \left\{	\begin{array}{ll}
		\displaystyle \frac{ (m_\gamma - \Lambda) \hbar }{ \rho }	\equiv \frac{  \ell \, \hbar} {\rho}
		&\qquad \text{canonical},	\\[2ex]
			\hbar \kappa 
			\displaystyle  \frac{ J_{m_\gamma}(\kappa \rho) }{ J_{m_\gamma-\Lambda}(\kappa \rho) }
			& \qquad \text{Belinfante}	.
			\end{array}	\right.
\end{align}

%%%%%%%%%%%%%%
%\begin{figure}[htbp]
%\begin{center}
%\includegraphics[width = 35 mm]{Figs_momcon2/CaRotor}
%\caption{A two-ion Calcium rotor.}
%\label{fig:rotor}
%\end{center}
%\end{figure}
%%%%%%%%%%%%%%

For a two $^{40}$Ca$^+$ ion rotor, Fig.~\ref{fig:twcylinder}(b), choose an atomic transition such as $4s_{1/2}  \to 4p_{3/2}$ or $4p_{1/2}$, where the excited state is not metastable but has a fast spontaneous decay.  The situation is analogous to laser cooling~\cite{1975OptCo..13...68H}: the spontaneous decay is isotropic so statistically there is no momentum kick in the decay, but the excitation always involves a momentum kick in the same azimuthal direction.  Shine the twisted photon beam so that its vortex line is perpendicular to the plane of the rotor and passes through its center.  If the exciting laser is strong enough to quickly excite the ground state ion, the ion will receive one momentum kick per lifetime of the excited state $T$.  This will give a force $dp/dt$,  a torque $\tau$, and for moment of inertia $I$, an angular acceleration   
\begin{align}
\alpha = \frac{ \tau } {I} = \frac{ 2 \rho \braket{ p_{\phi} } / T }{ 2 M \rho^2 }
	= \frac{ \braket{ p_{\phi} }  }{ M \rho T  }	\,,
\end{align}
where $M$ is the mass of the Calcium ion.    The lifetimes are
$T ( \text{Ca}^+, 4p_{3/2}) = 6.924(0.019)$\,ns and
$T ( \text{Ca}^+, 4p_{1/2}) = 7.098(0.020)$\,ns~\cite{1993PhRvL..70.3213J}.

\begin{figure}[htbp]
\begin{center}
\includegraphics[width = 85 mm]{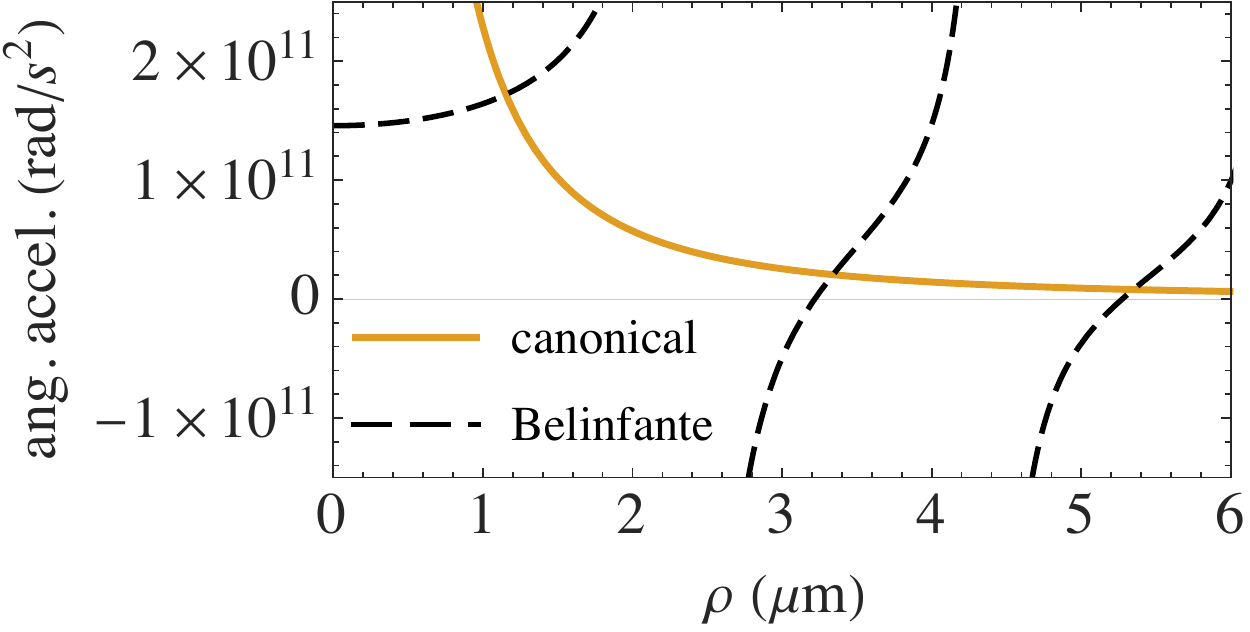}
\caption{Calculated angular acceleration for a two-ion Calcium rotor of varying radii, with further description in the text}
\label{fig:rotorangaccel}
\end{center}
\end{figure}

Fig.~\ref{fig:rotorangaccel} shows a plot of the angular acceleration vs. rotor radius for the $4p_{3/2}$ case; the $4p_{1/2}$ case is barely different.	

%%%%%%%%%%%%%%
%%%%%%%%%%%%%%

%%%%%%%%%%%%%%%%%%%%%%%%%%%%%%%%%%

\paragraph{Small particles off-axis in a twisted beam.}  Refs.~\cite{2002PhRvL..88e3601O,2003PhRvL..91i3602G} report measurements of the azimuthal kick received by small particles sitting at various distances or various impact parameters from the vortex line of a twisted photon beam.  (The azimuthal kick is given in terms of $\Omega_\text{revolution}$ in, say, Fig.~(2b) of~\cite{2003PhRvL..91i3602G}.)  The size of the kicks, given in terms of the azimuthal or transverse component of the momentum density is
\begin{align}
p_\phi =	\left\{		\begin{array}{l l}
		\epsilon_0 	\left[ \displaystyle{	\frac{ \omega\ell }{ \rho } } |u|^2  - \frac{1}{2 } \omega \sigma 
		\frac{ \eth |u|^2 }{\eth \rho} 		\right]  & \qquad \text{Belinfante}	\\[2.5 ex]
		\displaystyle{	\frac{ \epsilon_0 \omega\ell }{ \rho } } |u|^2	& \qquad \text{canonical}
		\end{array}		\right.
\end{align}
The Belinfante expression is also given in the middle line of Eq.~(2) of~\cite{2003PhRvL..91i3602G}; the canonical expression differs in the absence of the derivative term.  The measurement reported were made at the peaks of the intensity distribution in the rings of the twisted beam wavefront.  The intensity distribution is proportional to $|u|^2$, so these are precisely the locations where the Belinfante and canonical predictions are the same.  

\begin{figure}[htbp]

\hfil \includegraphics[width = \columnwidth] {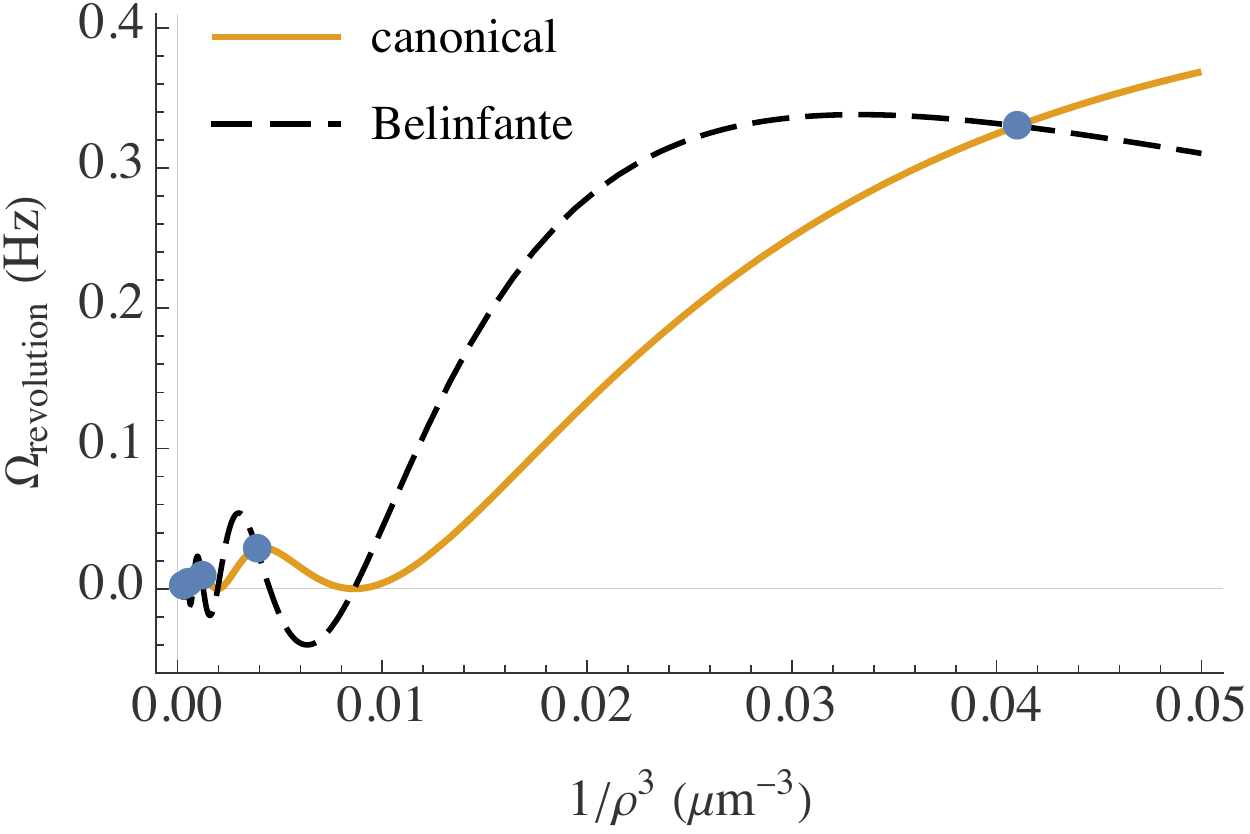}  \hfil

\vskip - 41 mm
\hskip 42.7 mm
\includegraphics[width = 0.5\columnwidth] {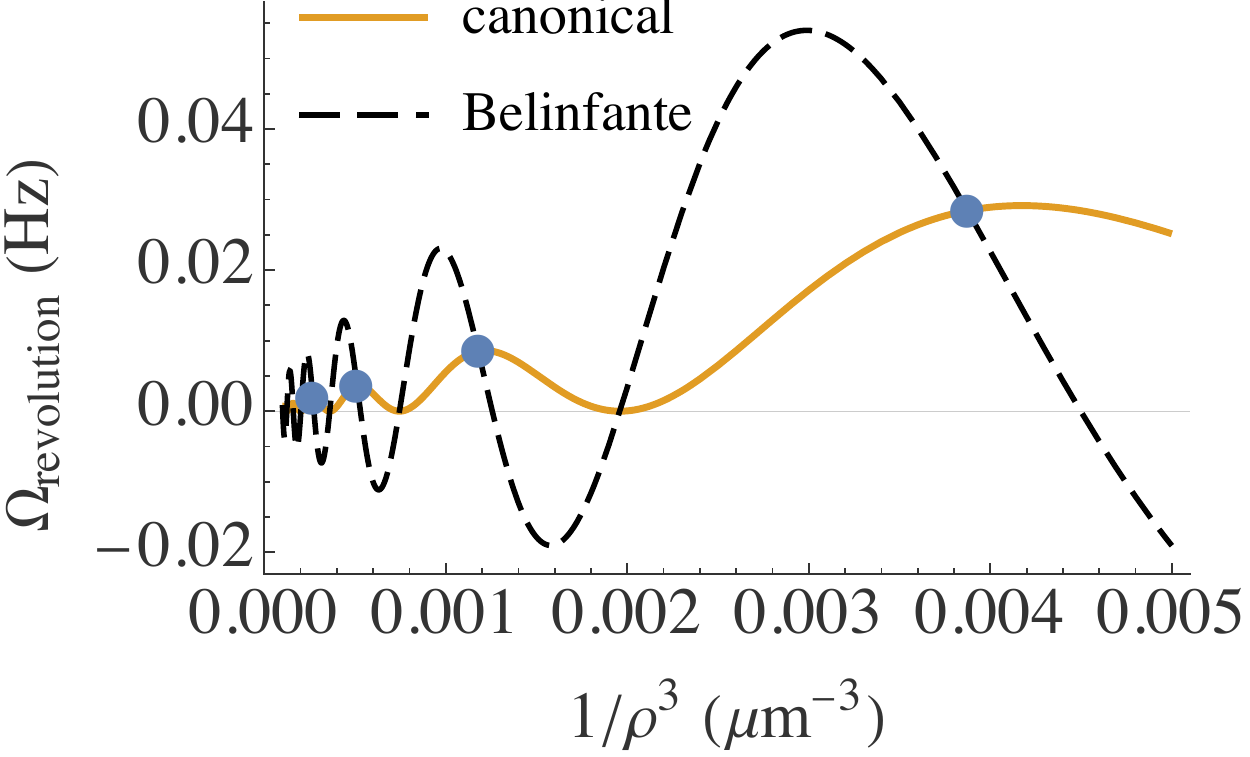}
\vskip 10 mm

\caption{Azimuthal kick given in terms of $\Omega_\text{revolution}$ vs. radius given as $1/\rho^3$ for a small particle in a twisted electromagnetic beam. Dots indicate where current measurements lie~\cite{2003PhRvL..91i3602G}.}
\label{fig:offaxiskicks}

\end{figure}

Fig.~\ref{fig:offaxiskicks} shows how the expectations from the two cases differ. The vertical axis shows the revolution frequency as a small particle at radius $\rho$ is kicked in a circular path about the vortex line.
The normalization depends on the power in the beam, and is chosen to match one the power settings in Fig.~(2b) of~\cite{2003PhRvL..91i3602G}. The dots show the
locations of the current measurements.

It would be worthwhile having measurements at other radii.  The zeros of intensity coincide with the zeros of canonical $\Omega_\text{revolution}$, so there are regions of good intensity where the canonical and Belinfante predictions differ significantly.

%%%%%%%%%%%%%%
%%%%%%%%%%%%%%

%%%%%%%%%%%%%%%%%%%%%%%%%%%%%%%%%%

%%%%%%%%%%%%%%%%%%%%%%%%%%%%%%%%%%

\paragraph{Radiation pressure from structured light.}

%%%%%%%%%%%%%%%%%%%%%%%%%%%%%%%%%%

Let us turn now to discussing the longitudinal component of the linear momentum density, and the radiation pressure forces engendered by that component.  The differences between the canonical and Belinfante predictions can be dramatic.  However, the dramatic differences are only in narrow regions and are sensitive to detail.  We will here work with exact Bessel beam expressions.  We also omit for now the Gaussian or other envelope.   The Bessel beam is built from photons that all have the same longitudinal momentum and same transverse momentum magnitude but varying azimuthal angles.  For the case where all the component photons have helicity $\Lambda$, the vector potential is~\cite{Jentschura:2010ap,Jentschura:2011ih,Afanasev:2013kaa}
{
\begin{align}
\label{eq:twistedwf}
&A^\mu_{\kappa m_\gamma k_z \Lambda}(\rho, \phi, z) = 				\nn\\
&\  -i \Lambda \, A \,
	e^{i(k_z z - \omega t + m_\gamma \phi )}	
	\bigg\{	e^{-i \Lambda \phi}  \cos^2\frac{\theta_k}{2} 	\,
	J_{m_\gamma-\Lambda}(\kappa\rho) \, \eta^\mu_\Lambda 
				\nn\\
&   + \frac{i}{\sqrt{2}}  \sin\theta_k	\,
	J_{m_\gamma}(\kappa\rho) \, \eta^\mu_0 
	-   e^{ i \Lambda \phi}  \sin^2\frac{\theta_k}{2} 	\,
	J_{m_\gamma+\Lambda}(\kappa\rho) \, \eta^\mu_{-\Lambda}
	\bigg\}
\end{align}}
where $k_z = k \cos\theta_k$ and $\kappa = k \sin\theta_k$.

The longitudinal components of momentum density %obtained from this potential 
are
\begin{align}
\braket{\mathcal P_{z}}_\text{can} &=\frac{\epsilon_0 \omega k_z A_0^{\,2} }{2}
	\bigg[ \cos^4 \frac{\theta_k}{2}  J_{m_\gamma-\Lambda}^2(\kappa \rho)	\nn\\
&	\hskip - 1 em
	+ \sin^4 \frac{\theta_k}{2}  J_{m_\gamma + \Lambda}^2(\kappa \rho)		
	+ \frac{1}{2} \sin^2 \theta_k \, J_{m_\gamma}^2(\kappa \rho)	\bigg]  \,,
	        \nn\\
\braket{\mathcal P_z}_\text{Bel}
 &= \frac{ \epsilon_0 \omega k A_0^{\,2} }{2}		\nn\\[1 ex]
 &	\hskip -2 em	\times 
	\left[  \cos^4 \frac{\theta_k}{2}  J^2_{m_\gamma-\Lambda}(\kappa \rho)
	- \sin^4 \frac{\theta_k}{2}  J^2_{m_\gamma + \Lambda}(\kappa \rho)	\right].
\end{align}
A test object of cross section $\sigma$ absorbing this momentum density feels a force 
$\braket{F_z} = \sigma c  \braket{ \mathcal P_z }$.

The momentum expressions are paraxially the same, and are very close numerically over broad regions.  Paraxially the states have $\sigma_z = \Lambda$.    However, in the full expressions
$\braket{\mathcal P_z}_\text{can}$ can never be negative for these modes, while 
$\braket{\mathcal P_z}_\text{Bel}$ is negative at and near radii $\rho$ where coefficient of the usually dominant $\cos^4(\theta_k/2)$ term becomes zero.

%%%%%%%%
\begin{figure}[t]
\begin{center}
\centerline{ (a) } \smallskip

\includegraphics[width = 0.98 \columnwidth]{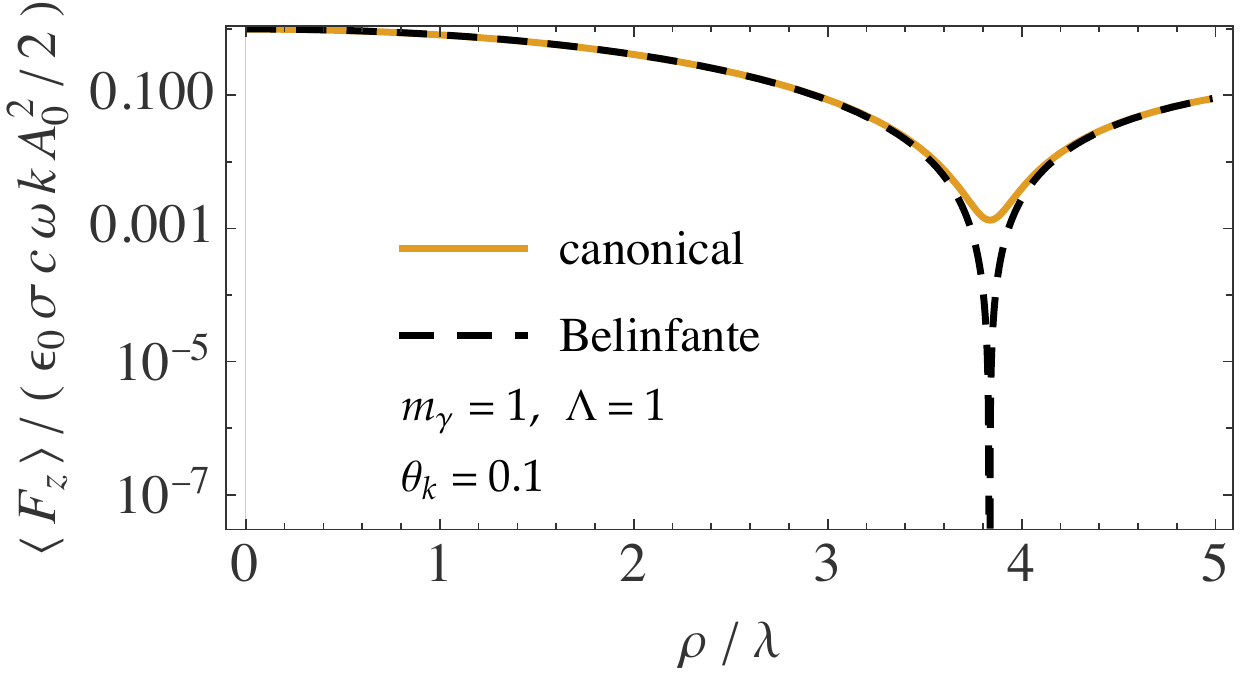}

\vskip 2 mm

\centerline{ (b) } \smallskip
\includegraphics[width = 0.98 \columnwidth]{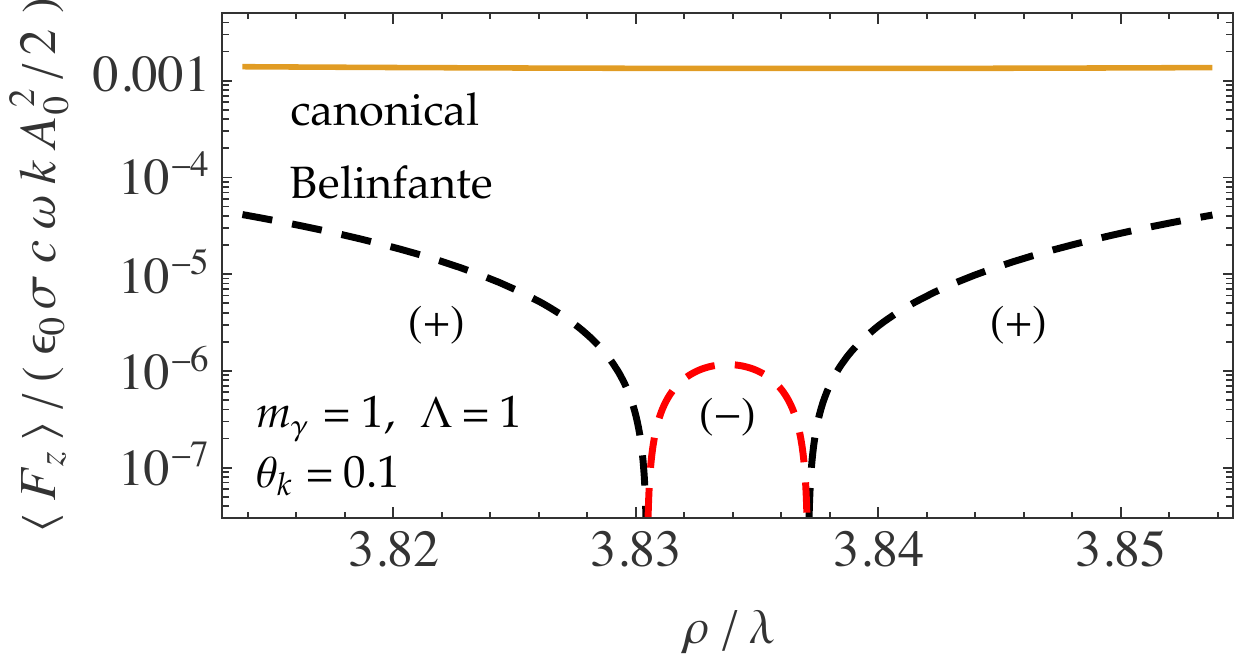}
\caption{Force from a twisted light beam on a small dielectric particle % of cross section $\sigma$, 
at distance $\rho$ from the photon's vortex line, for parameters stated on the plots.  Panel (a) gives a broad view and panel (b) gives a detail view.  %Parameters are $m_\gamma = 2$, $\Lambda =1$, and $\theta_k = 0.1$ radian. 
%The beam has total angular momentum $m_\gamma = 2$, helicity $\Lambda =1$,  pitch angle $\theta_k = 0.1$ radian, angular frequency $\omega$, and $A_0$ normalizing its strength.  
The canonical expression (solid gold) and Poynting vector expression (dashed, black when positive, red when negative) %for the momentum density of the beam 
give similar results except in regions near force minima. %where the Poynting vector for the twisted beam gets very small or in places, 
Note especially the red dashed curve in panel (b), where the Belinfante force gets negative, i.e., points opposite to the propagation direction of the beam.
}
\label{fig:linear2}
\end{center}
\end{figure}
%%%%%%%%

Fig.~\ref{fig:linear2} shows, for a selected $m_\gamma$, $\Lambda$, and $\theta_k$, the longitudinal force vs. distance from the vortex line for the two cases on a small test particle of cross section $\sigma$ that fully absorbs the beam that strikes it.  The results are nearly the same for long stretches of $\rho$, but the difference near the force minimum is dramatic.  Panel (b) focuses on a narrow region to emphasize the difference. The canonical case continues pushing in the propagation direction, but in the Belinfante case the radiation pressure becomes a tractor beam at these locations, that is, it pulls toward the source rather than pushes away.

%%%%%%%%%%%%%%%%%%%%%%%%%%%%%%%%%%

\paragraph{Summary.}

%%%%%%%%%%%%%%%%%%%%%%%%%%%%%%%%%%

The canonical and the symmetric or Belinfante forms of the electromagnetic energy-momentum tensor give identical results for integrated quantities such as the total momentum or total angular momentum of the field.  However, they differ point by point in space, and this matters for calculating the force or torque of an electromagnetic wave on a  test object of finite size.  One requires light with a structured wave front in order to see the differences, and we have worked out  examples using Bessel-Gauss beams of twisted photons.  In certain regions the differences are dramatic, including tractor beam effects - also noticed in Ref.\cite{Novitsky:07} - and counter-rotating torques predicted when using forces or torques derived from the symmetric momentum tensor.  The dramatic contrasts in the force lie in limited spatial regions and are sensitive to details of the beam preparation. The dramatic torque differences, however,  are robust and exist over broad spatial regions and could well be confirmed or denied experimentally using ringlike or end-weighted rotor %[]
%~\cite{2019PhRvL.123m3202U} 
test objects.    Numerical results suggest that the generated spin rate differences could be observable on micron sized objects, using available twisted photon beams.

%%%%%%%%%%%%%%%%%%%%%%%%%%%%%%%%%%

\paragraph{Acknowledgements.}
We thank Elliot Leader for inspiring conversations.
A.A.~thanks the US Army Research Office Grant W911NF-19-1-0022 for support and C.E.C.~thanks the National Science Foundation (USA) for support under grant PHY-1812326. A. M. thanks the SERB-POWER Fellowship, Department of Science and Technology, Govt. of India  for support.

%%%%%%%%%%%%%%%%%%%%%%%%%%%%%%%%%%

\bibliography{momcon}

\begin{thebibliography}{31}%
\makeatletter
\providecommand \@ifxundefined [1]{%
 \@ifx{#1\undefined}
}%
\providecommand \@ifnum [1]{%
 \ifnum #1\expandafter \@firstoftwo
 \else \expandafter \@secondoftwo
 \fi
}%
\providecommand \@ifx [1]{%
 \ifx #1\expandafter \@firstoftwo
 \else \expandafter \@secondoftwo
 \fi
}%
\providecommand \natexlab [1]{#1}%
\providecommand \enquote  [1]{``#1''}%
\providecommand \bibnamefont  [1]{#1}%
\providecommand \bibfnamefont [1]{#1}%
\providecommand \citenamefont [1]{#1}%
\providecommand \href@noop [0]{\@secondoftwo}%
\providecommand \href [0]{\begingroup \@sanitize@url \@href}%
\providecommand \@href[1]{\@@startlink{#1}\@@href}%
\providecommand \@@href[1]{\endgroup#1\@@endlink}%
\providecommand \@sanitize@url [0]{\catcode `\\12\catcode `\$12\catcode
  `\&12\catcode `\#12\catcode `\^12\catcode `\_12\catcode `\%12\relax}%
\providecommand \@@startlink[1]{}%
\providecommand \@@endlink[0]{}%
\providecommand \url  [0]{\begingroup\@sanitize@url \@url }%
\providecommand \@url [1]{\endgroup\@href {#1}{\urlprefix }}%
\providecommand \urlprefix  [0]{URL }%
\providecommand \Eprint [0]{\href }%
\providecommand \doibase [0]{http://dx.doi.org/}%
\providecommand \selectlanguage [0]{\@gobble}%
\providecommand \bibinfo  [0]{\@secondoftwo}%
\providecommand \bibfield  [0]{\@secondoftwo}%
\providecommand \translation [1]{[#1]}%
\providecommand \BibitemOpen [0]{}%
\providecommand \bibitemStop [0]{}%
\providecommand \bibitemNoStop [0]{.\EOS\space}%
\providecommand \EOS [0]{\spacefactor3000\relax}%
\providecommand \BibitemShut  [1]{\csname bibitem#1\endcsname}%
\let\auto@bib@innerbib\@empty
%</preamble>
\bibitem [{\citenamefont {{Yao}}\ and\ \citenamefont
  {{Padgett}}(2011)}]{2011AdOP....3..161Y}%
  \BibitemOpen
  \bibfield  {author} {\bibinfo {author} {\bibfnamefont {A.~M.}\ \bibnamefont
  {{Yao}}}\ and\ \bibinfo {author} {\bibfnamefont {M.~J.}\ \bibnamefont
  {{Padgett}}},\ }\href {\doibase 10.1364/AOP.3.000161} {\bibfield  {journal}
  {\bibinfo  {journal} {Advances in Optics and Photonics}\ }\textbf {\bibinfo
  {volume} {3}},\ \bibinfo {pages} {161} (\bibinfo {year} {2011})}\BibitemShut
  {NoStop}%
\bibitem [{\citenamefont {Bliokh}\ and\ \citenamefont
  {Nori}(2015)}]{Bliokh:2015doa}%
  \BibitemOpen
  \bibfield  {author} {\bibinfo {author} {\bibfnamefont {K.~Y.}\ \bibnamefont
  {Bliokh}}\ and\ \bibinfo {author} {\bibfnamefont {F.}~\bibnamefont {Nori}},\
  }\href {\doibase 10.1016/j.physrep.2015.06.003} {\bibfield  {journal}
  {\bibinfo  {journal} {Phys. Rept.}\ }\textbf {\bibinfo {volume} {592}},\
  \bibinfo {pages} {1} (\bibinfo {year} {2015})},\ \Eprint
  {http://arxiv.org/abs/1504.03113} {arXiv:1504.03113 [physics.optics]}
  \BibitemShut {NoStop}%
\bibitem [{\citenamefont {{Berry}}(2013)}]{2013EJPh...34.1337B}%
  \BibitemOpen
  \bibfield  {author} {\bibinfo {author} {\bibfnamefont {M.~V.}\ \bibnamefont
  {{Berry}}},\ }\href {\doibase 10.1088/0143-0807/34/6/1337} {\bibfield
  {journal} {\bibinfo  {journal} {European Journal of Physics}\ }\textbf
  {\bibinfo {volume} {34}},\ \bibinfo {eid} {1337} (\bibinfo {year}
  {2013})}\BibitemShut {NoStop}%
\bibitem [{\citenamefont {{Albaladejo}}\ \emph {et~al.}(2009)\citenamefont
  {{Albaladejo}}, \citenamefont {{Marqu{\'e}s}}, \citenamefont {{Laroche}},\
  and\ \citenamefont {{S{\'a}enz}}}]{2009PhRvL.102k3602A}%
  \BibitemOpen
  \bibfield  {author} {\bibinfo {author} {\bibfnamefont {S.}~\bibnamefont
  {{Albaladejo}}}, \bibinfo {author} {\bibfnamefont {M.~I.}\ \bibnamefont
  {{Marqu{\'e}s}}}, \bibinfo {author} {\bibfnamefont {M.}~\bibnamefont
  {{Laroche}}}, \ and\ \bibinfo {author} {\bibfnamefont {J.~J.}\ \bibnamefont
  {{S{\'a}enz}}},\ }\href {\doibase 10.1103/PhysRevLett.102.113602} {\bibfield
  {journal} {\bibinfo  {journal} {\prl}\ }\textbf {\bibinfo {volume} {102}},\
  \bibinfo {eid} {113602} (\bibinfo {year} {2009})}\BibitemShut {NoStop}%
\bibitem [{\citenamefont {{Huard}}\ and\ \citenamefont
  {{Imbert}}(1978)}]{1978OptCo..24..185H}%
  \BibitemOpen
  \bibfield  {author} {\bibinfo {author} {\bibfnamefont {S.}~\bibnamefont
  {{Huard}}}\ and\ \bibinfo {author} {\bibfnamefont {C.}~\bibnamefont
  {{Imbert}}},\ }\href {\doibase 10.1016/0030-4018(78)90115-3} {\bibfield
  {journal} {\bibinfo  {journal} {Optics Communications}\ }\textbf {\bibinfo
  {volume} {24}},\ \bibinfo {pages} {185} (\bibinfo {year} {1978})}\BibitemShut
  {NoStop}%
\bibitem [{\citenamefont {{Wang}}\ and\ \citenamefont
  {{Chen}}(2019)}]{2019PhRvA..99f3832W}%
  \BibitemOpen
  \bibfield  {author} {\bibinfo {author} {\bibfnamefont {Z.-L.}\ \bibnamefont
  {{Wang}}}\ and\ \bibinfo {author} {\bibfnamefont {X.-S.}\ \bibnamefont
  {{Chen}}},\ }\href {\doibase 10.1103/PhysRevA.99.063832} {\bibfield
  {journal} {\bibinfo  {journal} {\pra}\ }\textbf {\bibinfo {volume} {99}},\
  \bibinfo {eid} {063832} (\bibinfo {year} {2019})},\ \Eprint
  {http://arxiv.org/abs/1804.09013} {arXiv:1804.09013 [physics.optics]}
  \BibitemShut {NoStop}%
\bibitem [{\citenamefont {{Ornigotti}}\ and\ \citenamefont
  {{Aiello}}(2014)}]{2014OExpr..22.6586O}%
  \BibitemOpen
  \bibfield  {author} {\bibinfo {author} {\bibfnamefont {M.}~\bibnamefont
  {{Ornigotti}}}\ and\ \bibinfo {author} {\bibfnamefont {A.}~\bibnamefont
  {{Aiello}}},\ }\href {\doibase 10.1364/OE.22.006586} {\bibfield  {journal}
  {\bibinfo  {journal} {Optics Express}\ }\textbf {\bibinfo {volume} {22}},\
  \bibinfo {pages} {6586} (\bibinfo {year} {2014})},\ \Eprint
  {http://arxiv.org/abs/1304.5018} {arXiv:1304.5018 [physics.optics]}
  \BibitemShut {NoStop}%
\bibitem [{\citenamefont {{O'Neil}}\ \emph {et~al.}(2002)\citenamefont
  {{O'Neil}}, \citenamefont {{MacVicar}}, \citenamefont {{Allen}},\ and\
  \citenamefont {{Padgett}}}]{2002PhRvL..88e3601O}%
  \BibitemOpen
  \bibfield  {author} {\bibinfo {author} {\bibfnamefont {A.~T.}\ \bibnamefont
  {{O'Neil}}}, \bibinfo {author} {\bibfnamefont {I.}~\bibnamefont
  {{MacVicar}}}, \bibinfo {author} {\bibfnamefont {L.}~\bibnamefont {{Allen}}},
  \ and\ \bibinfo {author} {\bibfnamefont {M.~J.}\ \bibnamefont {{Padgett}}},\
  }\href {\doibase 10.1103/PhysRevLett.88.053601} {\bibfield  {journal}
  {\bibinfo  {journal} {\prl}\ }\textbf {\bibinfo {volume} {88}},\ \bibinfo
  {eid} {053601} (\bibinfo {year} {2002})}\BibitemShut {NoStop}%
\bibitem [{\citenamefont {{Garc{\'e}s-Ch{\'a}vez}}\ \emph
  {et~al.}(2003)\citenamefont {{Garc{\'e}s-Ch{\'a}vez}}, \citenamefont
  {{McGloin}}, \citenamefont {{Padgett}}, \citenamefont {{Dultz}},
  \citenamefont {{Schmitzer}},\ and\ \citenamefont
  {{Dholakia}}}]{2003PhRvL..91i3602G}%
  \BibitemOpen
  \bibfield  {author} {\bibinfo {author} {\bibfnamefont {V.}~\bibnamefont
  {{Garc{\'e}s-Ch{\'a}vez}}}, \bibinfo {author} {\bibfnamefont
  {D.}~\bibnamefont {{McGloin}}}, \bibinfo {author} {\bibfnamefont {M.~J.}\
  \bibnamefont {{Padgett}}}, \bibinfo {author} {\bibfnamefont {W.}~\bibnamefont
  {{Dultz}}}, \bibinfo {author} {\bibfnamefont {H.}~\bibnamefont
  {{Schmitzer}}}, \ and\ \bibinfo {author} {\bibfnamefont {K.}~\bibnamefont
  {{Dholakia}}},\ }\href {\doibase 10.1103/PhysRevLett.91.093602} {\bibfield
  {journal} {\bibinfo  {journal} {\prl}\ }\textbf {\bibinfo {volume} {91}},\
  \bibinfo {eid} {093602} (\bibinfo {year} {2003})}\BibitemShut {NoStop}%
\bibitem [{\citenamefont {{He}}\ \emph {et~al.}(1995)\citenamefont {{He}},
  \citenamefont {{Friese}}, \citenamefont {{Heckenberg}},\ and\ \citenamefont
  {{Rubinsztein-Dunlop}}}]{1995PhRvL..75..826H}%
  \BibitemOpen
  \bibfield  {author} {\bibinfo {author} {\bibfnamefont {H.}~\bibnamefont
  {{He}}}, \bibinfo {author} {\bibfnamefont {M.~E.~J.}\ \bibnamefont
  {{Friese}}}, \bibinfo {author} {\bibfnamefont {N.~R.}\ \bibnamefont
  {{Heckenberg}}}, \ and\ \bibinfo {author} {\bibfnamefont {H.}~\bibnamefont
  {{Rubinsztein-Dunlop}}},\ }\href {\doibase 10.1103/PhysRevLett.75.826}
  {\bibfield  {journal} {\bibinfo  {journal} {\prl}\ }\textbf {\bibinfo
  {volume} {75}},\ \bibinfo {pages} {826} (\bibinfo {year} {1995})}\BibitemShut
  {NoStop}%
\bibitem [{\citenamefont {{Friese}}\ \emph {et~al.}(1996)\citenamefont
  {{Friese}}, \citenamefont {{Enger}}, \citenamefont {{Rubinsztein-Dunlop}},\
  and\ \citenamefont {{Heckenberg}}}]{1996PhRvA..54.1593F}%
  \BibitemOpen
  \bibfield  {author} {\bibinfo {author} {\bibfnamefont {M.~E.~J.}\
  \bibnamefont {{Friese}}}, \bibinfo {author} {\bibfnamefont {J.}~\bibnamefont
  {{Enger}}}, \bibinfo {author} {\bibfnamefont {H.}~\bibnamefont
  {{Rubinsztein-Dunlop}}}, \ and\ \bibinfo {author} {\bibfnamefont {N.~R.}\
  \bibnamefont {{Heckenberg}}},\ }\href {\doibase 10.1103/PhysRevA.54.1593}
  {\bibfield  {journal} {\bibinfo  {journal} {\pra}\ }\textbf {\bibinfo
  {volume} {54}},\ \bibinfo {pages} {1593} (\bibinfo {year}
  {1996})}\BibitemShut {NoStop}%
\bibitem [{\citenamefont {Cirac}\ and\ \citenamefont
  {Zoller}(1995)}]{PhysRevLett.74.4091}%
  \BibitemOpen
  \bibfield  {author} {\bibinfo {author} {\bibfnamefont {J.~I.}\ \bibnamefont
  {Cirac}}\ and\ \bibinfo {author} {\bibfnamefont {P.}~\bibnamefont {Zoller}},\
  }\href {\doibase 10.1103/PhysRevLett.74.4091} {\bibfield  {journal} {\bibinfo
   {journal} {Phys. Rev. Lett.}\ }\textbf {\bibinfo {volume} {74}},\ \bibinfo
  {pages} {4091} (\bibinfo {year} {1995})}\BibitemShut {NoStop}%
\bibitem [{\citenamefont {{Schmiegelow}}\ \emph {et~al.}(2016)\citenamefont
  {{Schmiegelow}}, \citenamefont {{Schulz}}, \citenamefont {{Kaufmann}},
  \citenamefont {{Ruster}}, \citenamefont {{Poschinger}},\ and\ \citenamefont
  {{Schmidt-Kaler}}}]{2016NatCo...712998S}%
  \BibitemOpen
  \bibfield  {author} {\bibinfo {author} {\bibfnamefont {C.~T.}\ \bibnamefont
  {{Schmiegelow}}}, \bibinfo {author} {\bibfnamefont {J.}~\bibnamefont
  {{Schulz}}}, \bibinfo {author} {\bibfnamefont {H.}~\bibnamefont
  {{Kaufmann}}}, \bibinfo {author} {\bibfnamefont {T.}~\bibnamefont
  {{Ruster}}}, \bibinfo {author} {\bibfnamefont {U.~G.}\ \bibnamefont
  {{Poschinger}}}, \ and\ \bibinfo {author} {\bibfnamefont {F.}~\bibnamefont
  {{Schmidt-Kaler}}},\ }\href {\doibase 10.1038/ncomms12998} {\bibfield
  {journal} {\bibinfo  {journal} {Nature Communications}\ }\textbf {\bibinfo
  {volume} {7}},\ \bibinfo {eid} {12998} (\bibinfo {year} {2016})}\BibitemShut
  {NoStop}%
\bibitem [{\citenamefont {{Belinfante}}(1940)}]{1940Phy.....7..449B}%
  \BibitemOpen
  \bibfield  {author} {\bibinfo {author} {\bibfnamefont {F.~J.}\ \bibnamefont
  {{Belinfante}}},\ }\href {\doibase 10.1016/S0031-8914(40)90091-X} {\bibfield
  {journal} {\bibinfo  {journal} {Physica}\ }\textbf {\bibinfo {volume} {7}},\
  \bibinfo {pages} {449} (\bibinfo {year} {1940})}\BibitemShut {NoStop}%
\bibitem [{\citenamefont {Rosenfeld}(1940)}]{rosenfeld1940energy}%
  \BibitemOpen
  \bibfield  {author} {\bibinfo {author} {\bibfnamefont {L.}~\bibnamefont
  {Rosenfeld}},\ }\href@noop {} {\bibfield  {journal} {\bibinfo  {journal}
  {M{\'e}m. Acad. Roy. Belg.}\ }\textbf {\bibinfo {volume} {18}},\ \bibinfo
  {pages} {1} (\bibinfo {year} {1940})}\BibitemShut {NoStop}%
\bibitem [{\citenamefont {Leader}(2016)}]{Leader:2015vwa}%
  \BibitemOpen
  \bibfield  {author} {\bibinfo {author} {\bibfnamefont {E.}~\bibnamefont
  {Leader}},\ }\href {\doibase 10.1016/j.physletb.2016.03.023} {\bibfield
  {journal} {\bibinfo  {journal} {Phys. Lett. B}\ }\textbf {\bibinfo {volume}
  {756}},\ \bibinfo {pages} {303} (\bibinfo {year} {2016})},\ \Eprint
  {http://arxiv.org/abs/1510.03293} {arXiv:1510.03293 [hep-ph]} \BibitemShut
  {NoStop}%
\bibitem [{\citenamefont {Leader}(2018)}]{Leader:2017htb}%
  \BibitemOpen
  \bibfield  {author} {\bibinfo {author} {\bibfnamefont {E.}~\bibnamefont
  {Leader}},\ }\href {\doibase 10.1016/j.physletb.2018.02.029} {\bibfield
  {journal} {\bibinfo  {journal} {Phys. Lett. B}\ }\textbf {\bibinfo {volume}
  {779}},\ \bibinfo {pages} {385} (\bibinfo {year} {2018})},\ \Eprint
  {http://arxiv.org/abs/1710.03099} {arXiv:1710.03099 [physics.optics]}
  \BibitemShut {NoStop}%
\bibitem [{\citenamefont {{Barnett}}\ and\ \citenamefont
  {{Berry}}(2013)}]{2013JOpt...15l5701B}%
  \BibitemOpen
  \bibfield  {author} {\bibinfo {author} {\bibfnamefont {S.~M.}\ \bibnamefont
  {{Barnett}}}\ and\ \bibinfo {author} {\bibfnamefont {M.~V.}\ \bibnamefont
  {{Berry}}},\ }\href {\doibase 10.1088/2040-8978/15/12/125701} {\bibfield
  {journal} {\bibinfo  {journal} {Journal of Optics}\ }\textbf {\bibinfo
  {volume} {15}},\ \bibinfo {eid} {125701} (\bibinfo {year}
  {2013})}\BibitemShut {NoStop}%
\bibitem [{\citenamefont {Afanasev}\ \emph {et~al.}(2021)\citenamefont
  {Afanasev}, \citenamefont {Carlson},\ and\ \citenamefont
  {Mukherjee}}]{Afanasev_21}%
  \BibitemOpen
  \bibfield  {author} {\bibinfo {author} {\bibfnamefont {A.}~\bibnamefont
  {Afanasev}}, \bibinfo {author} {\bibfnamefont {C.~E.}\ \bibnamefont
  {Carlson}}, \ and\ \bibinfo {author} {\bibfnamefont {A.}~\bibnamefont
  {Mukherjee}},\ }\href {\doibase 10.1103/PhysRevResearch.3.023097} {\bibfield
  {journal} {\bibinfo  {journal} {Phys. Rev. Research}\ }\textbf {\bibinfo
  {volume} {3}},\ \bibinfo {pages} {023097} (\bibinfo {year}
  {2021})}\BibitemShut {NoStop}%
\bibitem [{\citenamefont {Ivanov}\ \emph {et~al.}(2022)\citenamefont {Ivanov},
  \citenamefont {Liu},\ and\ \citenamefont {Zhang}}]{Ivanov_22}%
  \BibitemOpen
  \bibfield  {author} {\bibinfo {author} {\bibfnamefont {I.~P.}\ \bibnamefont
  {Ivanov}}, \bibinfo {author} {\bibfnamefont {B.}~\bibnamefont {Liu}}, \ and\
  \bibinfo {author} {\bibfnamefont {P.}~\bibnamefont {Zhang}},\ }\href
  {\doibase 10.1103/PhysRevA.105.013522} {\bibfield  {journal} {\bibinfo
  {journal} {Phys. Rev. A}\ }\textbf {\bibinfo {volume} {105}},\ \bibinfo
  {pages} {013522} (\bibinfo {year} {2022})}\BibitemShut {NoStop}%
\bibitem [{\citenamefont {Leader}\ and\ \citenamefont
  {Lorc\`e}(2014)}]{Leader:2013jra}%
  \BibitemOpen
  \bibfield  {author} {\bibinfo {author} {\bibfnamefont {E.}~\bibnamefont
  {Leader}}\ and\ \bibinfo {author} {\bibfnamefont {C.}~\bibnamefont
  {Lorc\`e}},\ }\href {\doibase 10.1016/j.physrep.2014.02.010} {\bibfield
  {journal} {\bibinfo  {journal} {Phys. Rept.}\ }\textbf {\bibinfo {volume}
  {541}},\ \bibinfo {pages} {163} (\bibinfo {year} {2014})},\ \Eprint
  {http://arxiv.org/abs/1309.4235} {arXiv:1309.4235 [hep-ph]} \BibitemShut
  {NoStop}%
\bibitem [{\citenamefont {Jauch}\ and\ \citenamefont
  {Rohrlich}(1976)}]{Jauch:1976ava}%
  \BibitemOpen
  \bibfield  {author} {\bibinfo {author} {\bibfnamefont {J.}~\bibnamefont
  {Jauch}}\ and\ \bibinfo {author} {\bibfnamefont {F.}~\bibnamefont
  {Rohrlich}},\ }\href {\doibase 10.1007/978-3-642-80951-4} {\emph {\bibinfo
  {title} {{The theory of photons and electrons. The relativistic quantum field
  theory of charged particles with spin one-half}}}},\ \bibinfo {edition}
  {2nd}\ ed.,\ Texts and Monographs in Physics\ (\bibinfo  {publisher}
  {Springer},\ \bibinfo {address} {Berlin},\ \bibinfo {year}
  {1976})\BibitemShut {NoStop}%
\bibitem [{\citenamefont {Bjorken}\ and\ \citenamefont
  {Drell}(1965)}]{Bjorken:1965zz}%
  \BibitemOpen
  \bibfield  {author} {\bibinfo {author} {\bibfnamefont {J.~D.}\ \bibnamefont
  {Bjorken}}\ and\ \bibinfo {author} {\bibfnamefont {S.~D.}\ \bibnamefont
  {Drell}},\ }\href@noop {} {\emph {\bibinfo {title} {{Relativistic Quantum
  Fields}}}},\ International Series In Pure and Applied Physics\ (\bibinfo
  {publisher} {McGraw-Hill},\ \bibinfo {address} {New York},\ \bibinfo {year}
  {1965})\BibitemShut {NoStop}%
\bibitem [{\citenamefont {{Afanasev}}\ \emph {et~al.}(2018)\citenamefont
  {{Afanasev}}, \citenamefont {{Carlson}}, \citenamefont {{Schmiegelow}},
  \citenamefont {{Schulz}}, \citenamefont {{Schmidt-Kaler}},\ and\
  \citenamefont {{Solyanik}}}]{2018NJPh...20b3032A}%
  \BibitemOpen
  \bibfield  {author} {\bibinfo {author} {\bibfnamefont {A.}~\bibnamefont
  {{Afanasev}}}, \bibinfo {author} {\bibfnamefont {C.~E.}\ \bibnamefont
  {{Carlson}}}, \bibinfo {author} {\bibfnamefont {C.~T.}\ \bibnamefont
  {{Schmiegelow}}}, \bibinfo {author} {\bibfnamefont {J.}~\bibnamefont
  {{Schulz}}}, \bibinfo {author} {\bibfnamefont {F.}~\bibnamefont
  {{Schmidt-Kaler}}}, \ and\ \bibinfo {author} {\bibfnamefont {M.}~\bibnamefont
  {{Solyanik}}},\ }\href {\doibase 10.1088/1367-2630/aaa63d} {\bibfield
  {journal} {\bibinfo  {journal} {New Journal of Physics}\ }\textbf {\bibinfo
  {volume} {20}},\ \bibinfo {eid} {023032} (\bibinfo {year} {2018})},\ \Eprint
  {http://arxiv.org/abs/1709.05571} {arXiv:1709.05571 [quant-ph]} \BibitemShut
  {NoStop}%
\bibitem [{\citenamefont {{Urban}}\ \emph {et~al.}(2019)\citenamefont
  {{Urban}}, \citenamefont {{Glikin}}, \citenamefont {{Mouradian}},
  \citenamefont {{Krimmel}}, \citenamefont {{Hemmerling}},\ and\ \citenamefont
  {{Haeffner}}}]{2019PhRvL.123m3202U}%
  \BibitemOpen
  \bibfield  {author} {\bibinfo {author} {\bibfnamefont {E.}~\bibnamefont
  {{Urban}}}, \bibinfo {author} {\bibfnamefont {N.}~\bibnamefont {{Glikin}}},
  \bibinfo {author} {\bibfnamefont {S.}~\bibnamefont {{Mouradian}}}, \bibinfo
  {author} {\bibfnamefont {K.}~\bibnamefont {{Krimmel}}}, \bibinfo {author}
  {\bibfnamefont {B.}~\bibnamefont {{Hemmerling}}}, \ and\ \bibinfo {author}
  {\bibfnamefont {H.}~\bibnamefont {{Haeffner}}},\ }\href {\doibase
  10.1103/PhysRevLett.123.133202} {\bibfield  {journal} {\bibinfo  {journal}
  {\prl}\ }\textbf {\bibinfo {volume} {123}},\ \bibinfo {eid} {133202}
  (\bibinfo {year} {2019})},\ \Eprint {http://arxiv.org/abs/1903.05763}
  {arXiv:1903.05763 [quant-ph]} \BibitemShut {NoStop}%
\bibitem [{\citenamefont {{Hansch}}\ and\ \citenamefont
  {{Schawlow}}(1975)}]{1975OptCo..13...68H}%
  \BibitemOpen
  \bibfield  {author} {\bibinfo {author} {\bibfnamefont {T.~W.}\ \bibnamefont
  {{Hansch}}}\ and\ \bibinfo {author} {\bibfnamefont {A.~L.}\ \bibnamefont
  {{Schawlow}}},\ }\href {\doibase 10.1016/0030-4018(75)90159-5} {\bibfield
  {journal} {\bibinfo  {journal} {Optics Communications}\ }\textbf {\bibinfo
  {volume} {13}},\ \bibinfo {pages} {68} (\bibinfo {year} {1975})}\BibitemShut
  {NoStop}%
\bibitem [{\citenamefont {{Jin}}\ and\ \citenamefont
  {{Church}}(1993)}]{1993PhRvL..70.3213J}%
  \BibitemOpen
  \bibfield  {author} {\bibinfo {author} {\bibfnamefont {J.}~\bibnamefont
  {{Jin}}}\ and\ \bibinfo {author} {\bibfnamefont {D.~A.}\ \bibnamefont
  {{Church}}},\ }\href {\doibase 10.1103/PhysRevLett.70.3213} {\bibfield
  {journal} {\bibinfo  {journal} {\prl}\ }\textbf {\bibinfo {volume} {70}},\
  \bibinfo {pages} {3213} (\bibinfo {year} {1993})}\BibitemShut {NoStop}%
\bibitem [{\citenamefont {Jentschura}\ and\ \citenamefont
  {Serbo}(2011{\natexlab{a}})}]{Jentschura:2010ap}%
  \BibitemOpen
  \bibfield  {author} {\bibinfo {author} {\bibfnamefont {U.}~\bibnamefont
  {Jentschura}}\ and\ \bibinfo {author} {\bibfnamefont {V.}~\bibnamefont
  {Serbo}},\ }\href {\doibase 10.1103/PhysRevLett.106.013001} {\bibfield
  {journal} {\bibinfo  {journal} {Phys. Rev. Lett.}\ }\textbf {\bibinfo
  {volume} {106}},\ \bibinfo {pages} {013001} (\bibinfo {year}
  {2011}{\natexlab{a}})},\ \Eprint {http://arxiv.org/abs/1008.4788}
  {arXiv:1008.4788 [physics.acc-ph]} \BibitemShut {NoStop}%
\bibitem [{\citenamefont {Jentschura}\ and\ \citenamefont
  {Serbo}(2011{\natexlab{b}})}]{Jentschura:2011ih}%
  \BibitemOpen
  \bibfield  {author} {\bibinfo {author} {\bibfnamefont {U.}~\bibnamefont
  {Jentschura}}\ and\ \bibinfo {author} {\bibfnamefont {V.}~\bibnamefont
  {Serbo}},\ }\href {\doibase 10.1140/epjc/s10052-011-1571-z} {\bibfield
  {journal} {\bibinfo  {journal} {Eur. Phys. J. C}\ }\textbf {\bibinfo {volume}
  {71}},\ \bibinfo {pages} {1571} (\bibinfo {year} {2011}{\natexlab{b}})},\
  \Eprint {http://arxiv.org/abs/1101.1206} {arXiv:1101.1206 [physics.acc-ph]}
  \BibitemShut {NoStop}%
\bibitem [{\citenamefont {Afanasev}\ \emph {et~al.}(2013)\citenamefont
  {Afanasev}, \citenamefont {Carlson},\ and\ \citenamefont
  {Mukherjee}}]{Afanasev:2013kaa}%
  \BibitemOpen
  \bibfield  {author} {\bibinfo {author} {\bibfnamefont {A.}~\bibnamefont
  {Afanasev}}, \bibinfo {author} {\bibfnamefont {C.~E.}\ \bibnamefont
  {Carlson}}, \ and\ \bibinfo {author} {\bibfnamefont {A.}~\bibnamefont
  {Mukherjee}},\ }\href {\doibase 10.1103/PhysRevA.88.033841} {\bibfield
  {journal} {\bibinfo  {journal} {Phys. Rev. A}\ }\textbf {\bibinfo {volume}
  {88}},\ \bibinfo {pages} {033841} (\bibinfo {year} {2013})},\ \Eprint
  {http://arxiv.org/abs/1305.3650} {arXiv:1305.3650 [quant-ph]} \BibitemShut
  {NoStop}%
\bibitem [{\citenamefont {Novitsky}\ and\ \citenamefont
  {Novitsky}(2007)}]{Novitsky:07}%
  \BibitemOpen
  \bibfield  {author} {\bibinfo {author} {\bibfnamefont {A.~V.}\ \bibnamefont
  {Novitsky}}\ and\ \bibinfo {author} {\bibfnamefont {D.~V.}\ \bibnamefont
  {Novitsky}},\ }\href {\doibase 10.1364/JOSAA.24.002844} {\bibfield  {journal}
  {\bibinfo  {journal} {J. Opt. Soc. Am. A}\ }\textbf {\bibinfo {volume}
  {24}},\ \bibinfo {pages} {2844} (\bibinfo {year} {2007})}\BibitemShut
  {NoStop}%
\end{thebibliography}%

\end{document}